# Large-scale road network partitioning: a deep learning method based on convolutional autoencoder model

**Pengfei Xu**
Graduate Student
Urban Mobility Institute,
Tongji University
4800 Cao'an Road, Shanghai 201804, China
Tel: +86-15255567973
Email: 2031406@tongji.edu.cn

**Weifeng Li**
Ph.D., Associate Researcher
Key Laboratory of Road and Traffic Engineering of the Ministry of Education,
College of Transportation Engineering,
Tongji University
4800 Cao'an Road, Shanghai 201804, China
Tel: +86-21-69583775
E-mail: liweifeng@tongji.edu.cn

**Chenjie Xu**
Graduate Student
Key Laboratory of Road and Traffic Engineering of the Ministry of Education,
College of Transportation Engineering,
Tongji University
4800 Cao'an Road, Shanghai 201804, China
Tel: +86-13917337931
E-mail: 2031344@tongji.edu.cn

**Jian Li (corresponding author)**
Ph.D., Associate Professor
Urban Mobility Institute,
College of Transportation Engineering,
Tongji University
4800 Cao'an Road, Shanghai 201804, China
Tel: +86-21-69583697
Email: jianli@tongji.edu.cn

Word Count: 5759 words + 2 table (250 words per table) = 6074 words

*Submitted in 2022.8.1*

*Pengfei Xu, Weifeng Li, and Jian Li*

## ABSTRACT

With the development of urbanization, the scale of urban road network continues to expand, especially in some Asian countries. Short-term traffic state prediction is one of the bases of traffic management and control. Constrained by the space-time cost of computation, the short-term traffic state prediction of large-scale urban road network is difficult. One way to solve this problem is to partition the whole network into multiple sub-networks to predict traffic state separately. In this study, a deep learning method is proposed for road network partitioning. The method mainly includes three steps. First, the daily speed series for roads are encoded into the matrix. Second, a convolutional autoencoder (AE) is built to extract series features and compress data. Third, the spatial hierarchical clustering method with adjacency relationship is applied in the road network. The proposed method was verified by the road network of Shenzhen which contains more than 5000 links. The results show that AE-hierarchical clustering distinguishes the tidal traffic characteristics and reflects the process of congestion propagation. Furthermore, two indicators are designed to make a quantitative comparison with spectral clustering: intra homogeneity increases by about 9% on average while inter heterogeneity about 9.5%. Compared with past methods, the time cost also decreases. The results may suggest ways to improve the management and control of urban road network in other metropolitan cities. The proposed method is expected to be extended to other problems related to large-scale networks.

**Keywords:** Road network partitioning, Urban traffic characteristics, Gram matrix, Autoencoder, Spatial hierarchical clustering

## 1. INTRODUCTION

Around the world, the scale of cities in many developing countries is experiencing an unprecedented expansion. Taking China as an example, from 1981 to 2017, the urban built-up area increased from 7438 square kilometers to 56225 square kilometers, an increase of 6.6 times; the population rose from 0.852 billion to 1.4 billion. Limited by the cost of computing resources, timely and precise traffic service in large-scale urban road network has gradually become a difficult problem. The key to providing these services lies in the accurate and prompt prediction of the road network traffic states.

There are two main solutions for above problems. One is to build an online simulation system for the city-level road network. However, this method is still constrained by the scale of the road network and the simulation tools, and the requirements for the hardware and software of the system are high. The other one is to partition the whole road network into several independent sub-networks for separate control (*1*). This method greatly reduces the difficulty of system operation, which is a widely accepted solution now. By road network partitioning, differentiated and refined control measures can be realized. Flexible control schemes can be adopted in more homogeneous sub-networks which means that independent traffic control strategies can be implemented within each subnetwork. Furthermore, partitioning may also improve the accuracy of the road network traffic state prediction model with graph structure and reduce the time cost of model training procedure by parallelization(*2*).

Most of the existing partitioning methods are graph-based methods based on topological geometry, which may be difficult to mine the characteristics of road traffic state in a deep level (*3–5*). Deep learning model has strong feature extraction ability and good adaptability to large-scale complex problems. It has shown exciting performance in traffic state prediction and many other problems. However, there are few applications of deep learning method in road network partitioning problems, and its potential remains to be tapped. Therefore, this study will use the deep learning method to extract the fluctuation characteristics of road traffic state as the basis of road network partitioning.

In this study, we propose a road network partitioning framework called AE- hierarchical clustering based on topological and operational characteristics using deep learning methods. The average speed series data of roads collected by floating cars in Shenzhen network are used for a case study. The framework first encodes time series to Gramian Angular Fields (GAF) (*6*). Next, a convolutional autoencoder (AE) is built both input and output of which are GAF. The autoencoder realizes data dimensionality reduction and denoising in the process of data reconstruction. Finally, we use encoder output as the similarity measure, apply spatial hierarchical clustering considering adjacency relationship in road network, and then we get multiple internally connected sub-networks. The proposed framework is compared with spectral clustering qualitatively and quantitatively.

Several features distinguish this study from previous ones.

First, the computer vision and representation learning method in deep learning are combined to extract the effective representation of road traffic state series, which shows the application potential of deep learning method in large-scale road network partitioning problems. The study expands the application of representational learning (*7*) to a certain extent.

Second, the topological characteristics of the road network are considered. The roads must be adjacent to be in the same sub-network so that each sub-network is completely connected. Compared with some other studies (*8*), our methods can obtain internal connected sub-network which has practical value for the implementation of perimeter or region control.

Third, the time cost of previous partitioning algorithms is generally high (*3*, *9*). For the methods proposed in this study, though the deep learning models usually need long-time training, the running speed after training is faster than other methods. It is sufficient to retrain the model periodically or when the road network topology changes significantly.

The remainder of this paper is organized as follows. In section 2, a brief overview of research related to this study is provided. In section 3, the case study area and dataset are described. In section 4, the methodology and each algorithmic step are explained in detail. In section 5, partition results of the proposed framework are presented and comparison with other approaches for the case study is conducted. In section 6, some discussions about our research are elaborated. Conclusions and directions for future study are presented in the last section.

## 2. LITERATURE REVIEW

Graph partition problem (GPP) is the basis of road network partitioning. The representative studies include spectral graph partition problems (SGPPs), graph cutting problems, etc. GPP aims to mine the internal relations and association patterns between data and provide valuable information for decision-makers. It has been successfully applied in many fields, including computer processing unit allocation (*10*), community detection (*11*) and image division quality evaluation (*12*). The current graph partition methods can be divided into two categories: algebra-based (*12–14*) and combinatorial optimization-based. Traditional algebraic methods face problems such as high space-time complexity and computational expense when dealing with large-scale problems. Therefore, some scholars considered combining heuristics with algebraic methods and proposed some combinatorial algorithms (*13*). These algorithms have better performance in solving efficiency. With the advent of the big data era, these algorithms are gradually facing bottlenecks. Using machine learning methods to improve the performance of evolutionary algorithms (*15*, *16*) has become frontier research in this field. However, the methods combined with machine learning mechanisms mainly focus on solving classic combinatorial optimization problems, and the application scenarios are still limited.

Some scholars modified the classic algorithm in the GGP and applied it to the road network partitioning problem, but there are some shortages. Dimitriou and Nikolaou (*8*) compared the effects of K-means and METIS[1] on the road network. The results of the two methods were poor. Etemadnia et al. (*17*) proposed two heuristic algorithms. The optimization objectives of the two algorithms are almost the same as N-cut. Ji and Geroliminis (*18*) designed a multi-step algorithm. Firstly, the N-cut algorithm is used to over divide the road network, and then a merging algorithm is designed to obtain the rough results. Finally, the boundary of each sub-network is adjusted to reduce the traffic state heterogeneity in the sub-network. There are two main shortcomings in applying graph partition algorithms directly to the road network. Firstly, these algorithms have high complexity, which leads to unbearable computational time and space in large-scale urban road network; Secondly, the objectives. of these algorithms cannot fulfill the need of road network partition. For example, the classic algorithm N-cut tends to cut the groups with the least contact, while road network partitioning prefers to divide the whole network into multiple sub-networks with high internal state homogeneity.

Based on above shortcomings, some studies proposed road network partitioning methods considering the characteristics of the actual transportation system. Saeedmanesh and Geroliminis (*3*) proposed a three-step clustering algorithm from the perspective of congestion propagation. An optimized three-step clustering algorithm was introduced in Saeedmanesh and Geroliminis (*9*), which added integer linear programming. The results show that there are obvious vehicle density differences between obtained sub-networks. Anwar et al. (*4*) proposed a hypergraph algorithm. The road network is abstracted as an undirected graph with roads as nodes and intersections as edges. Then, a hypergraph is constructed according to the vehicle density and spatial location of roads. Finally, an α-cut algorithm is used to cut the hypergraph. Compared with other algorithms, the hypergraph algorithm performs better. Another way of constructing hypergraphs was proposed in Anwar et al. (*5*) . To alleviate the disadvantage of high computational complexity, Anwar et al. (*19*) proposed an update method that updates the sub-networks on a small scale. Several studies in this field are summarized in Table 1.

In general, many studies have shown that it is not appropriate to directly apply graph partition algorithms to road network. Some studies have proposed various methods specified for road network, but they still face problems such as high data quality requirements, complex calculation process, and slow speed. Additionally, deep learning methods often appeared in traffic state prediction in network-related problems(*20–23*) while are rarely used in road network partitioning. Therefore, how to divide road network appropriately is still worthy of in-depth discussion.

---

[1] Graph partitioning software package developed by karypis laboratory in the USA.

**Table 1**
Some studies focusing on the road network partitioning problem

| Reference | Methods | Datasets[2] | Model | Advantages | Disadvantages |
|---|---|---|---|---|---|
| Dimitriou and Nikolaou (**8**) | GGP – based methods | Dataset 1 | METIS and K-means. | The model is simple and fast. | No obvious difference in the state of sub-networks. |
| Etemadnia et al. (**17**) | | Unspecified. | Two heuristic algorithms. | Maintain dominating traffic movement within sub-networks. | Weak homogeneity within each sub-network. |
| Ji and Geroliminis (**18**) | | | A three-step clustering algorithm. | The MFD[3] are obviously different among the sub-networks. | Focus on static partitioning and not involved dynamic partitioning. |
| Saeedmanesh and Geroliminis (**3**) | Congestion propagation – based methods | Dataset 5, 6 | A three-step clustering algorithm. | High homogeneity within each sub-network. | High data quality requirements (fully connected two-way road network) and high spatiotemporal complexity. |
| Saeedmanesh and Geroliminis (**9**) | | Dataset 6 | A mixed integer linear optimization algorithm. | | |
| Anwar et al. (**4**) | | Dataset 2, 3, 5 | A two-stage algorithm. | High homogeneity within each sub-network and high heterogeneity between sub-networks. | High spatiotemporal complexity (about 2000 seconds for one partition of dataset 2). |
| Anwar et al. (**5**) | | | | | |
| Anwar et al. (**19**) | Optimization methods | Dataset 2, 3, 4 | A two-layer algorithm. | Reduce the running cost of partitioning algorithms. | \ |
| Xu and Jacobsen (**24**) | | No case. | A novel partitioning schemes. | | The research object is not road network partitioning. |

## 3. STUDY AREA AND DATASET
### 3.1 Study area

   Shenzhen is selected as the study area in this research. Shenzhen is one of the most developed and urbanized cities in China. As the national center of economy, science, and technology, by the end of 2020,

---

[2] Dataset 1: 5735 links of Nicosia; Dataset 2: 17206 links of CBD Melbourne; Dataset 3: 53494 links of CBD Melbourne; Dataset 4: 7245 links of Melbourne; Dataset 5: 12305 links of Shenzhen; Dataset 6: 2.5 square mile area of Downtown San Francisco.

[3] Macro fundamental diagrams.

Shenzhen has a population of more than 17.56 million with a total area of 1997.47 square kilometers. Shenzhen has a busy road traffic system with a total length of 6124.2 kilometers of roads at all levels, undertaking the travel of more than 3.53 million vehicles, and its vehicles ownership ranks first in China. In addition to private transport modes, public transit also plays an important role in Shenzhen. According to statistical data of 2020, with 898 routes and 35809 vehicles, trips by the ground bus system reach 1.54 billion. The above statistics show that the road network of Shenzhen is very large in scale and has a number of private and public travel, so it is significant to take Shenzhen as a reference. Figure 1 shows the spatial distribution of 10 administrative districts of Shenzhen.

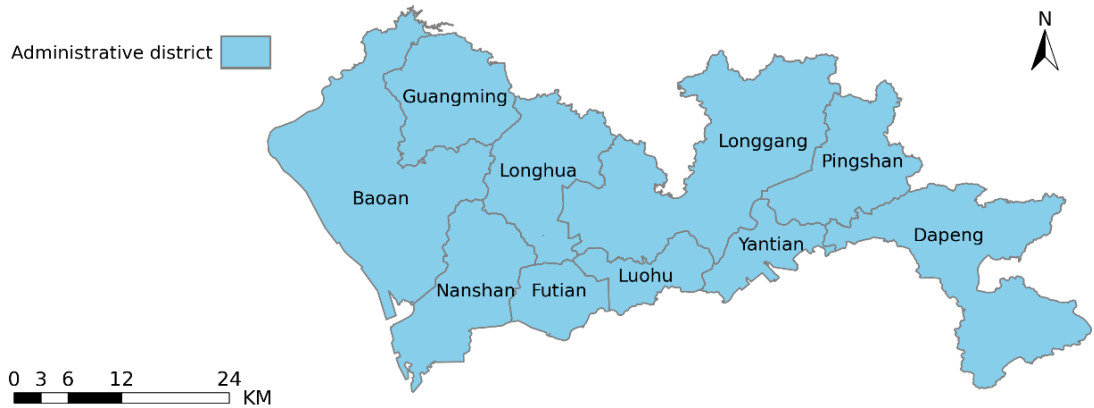

**Figure 1 Study area**

### 3.2 Dataset

The operation data used in this study are collected in 3945 roads, which are secondary roads and above, in Shenzhen road network by floating cars from 21 June to 25 June, 2021. The data attributes and examples are shown in Table 2. In the preprocess stage, due to the small number of floating cars in late night and early morning hours (from 22:00 to 7:00), there are many missing data in this period, and a Bayes-based matrix decomposition method (*25*) was used to fill in the missing records.

**TABLE 2 Explanation of case data**

| Attributes | Data Example | Explanation |
| :---: | :---: | :---: |
| **Date** | 20190909 | 20190909 means 9 September, 2019 |
| **Period** | 1 | 1 day is divided into 288 time intervals, each of which is 5 minutes long |
| **Road_id** | 3 | ID of road |
| **Speed** | 25.14 | Average speed of the road in current time period |
| **Sample_vehicles** | 20 | Number of floating cars used to calculate the average speed of the road section |

## 4.  METHODS

### 4.1 Encoding time series to images by Gramian Angular Field

Previous studies utilized the average speed of each road in different time periods (like morning peak and evening peak) to calculate the similarity to partition the road network, while this study is directly based on the traffic state time series of each road. The reasons for using traffic flow time series instead of the average value of specific time periods are: (i) It can better reflect the traffic state changing with time and prevent the disappearance of sequence fluctuation characteristics caused by taking the average value; (ii) A relatively stable partitioning can better coordinate with traffic management measures; (iii) It may be a trade-off between complexity and accuracy, despite the fact that the shorter the partition interval, the more accurate the results are, the traffic control system is under great pressure if the partition interval is excessively small, such as timely state switching.

To make better use of the success of deep learning in computer vision and speech recognition, we encode daily speed series as images, namely, Gramian Angular Filed (GAF) (Wang and Oates, 2015). The overall process of encoding time series into a matrix is shown in Figure 2, Given a time series $V = \{v_0, \; v_1, \ldots, v_n\}$, first we normalize all values by Equation 1 to make them fall in the interval [-1,1].

$$\widetilde{v_i} = \frac{(v_i - \max(V) + (v_i - \min(V)))}{\max(V) - \min(V)} \tag{1}$$

Second, the rescaled series can be represented in polar coordinate by calculating angular cosine and radius from value and time interval with the Equation 2.

$$\begin{cases} \emptyset = arccos(\widetilde{v_i}), -1 \leq \widetilde{v_i} \leq 1, \widetilde{v_i} \in \widetilde{V} \\ r = \dfrac{t_i}{N}, t_i \in N \end{cases} \tag{2}$$

In the formula, $t_i$ is the index of time interval, $\widetilde{V}$ is the normalized time series and N is a constant to adjust the span of the polar coordinate. As time increases, corresponding values have different angular points, spreading out like water waves on the spanning circles.

Third, the temporal correlation with different time intervals could be obtained by the trigonometric sum between each point. The GAF is defined with the equation below.

$$G = \begin{bmatrix} cos(\emptyset_1 + \emptyset_1) & \cdots & cos(\emptyset_1 + \emptyset_n) \\ \vdots & \ddots & \vdots \\ cos(\emptyset_n + \emptyset_1) & \cdots & cos(\emptyset_n + \emptyset_n) \end{bmatrix} = \widetilde{V}' \cdot \widetilde{V} - \sqrt{I - \widetilde{V}^2}' \sqrt{I - \widetilde{V}^2} \tag{3}$$

I is the unit vector [1,1,…,1]. The value in GAF reflects the correlation between different time intervals, and the diagonal value contains the original value information of the time series. Since the number of time intervals of a day is 288, the GAF matrix size is 288x288. One daily average speed sequence of a road corresponds to one GAF matrix. Encoding series to matrix allows us to use computer vision methods, such as Convolutional Neural Networks (CNN), to extract series fluctuation characteristics.

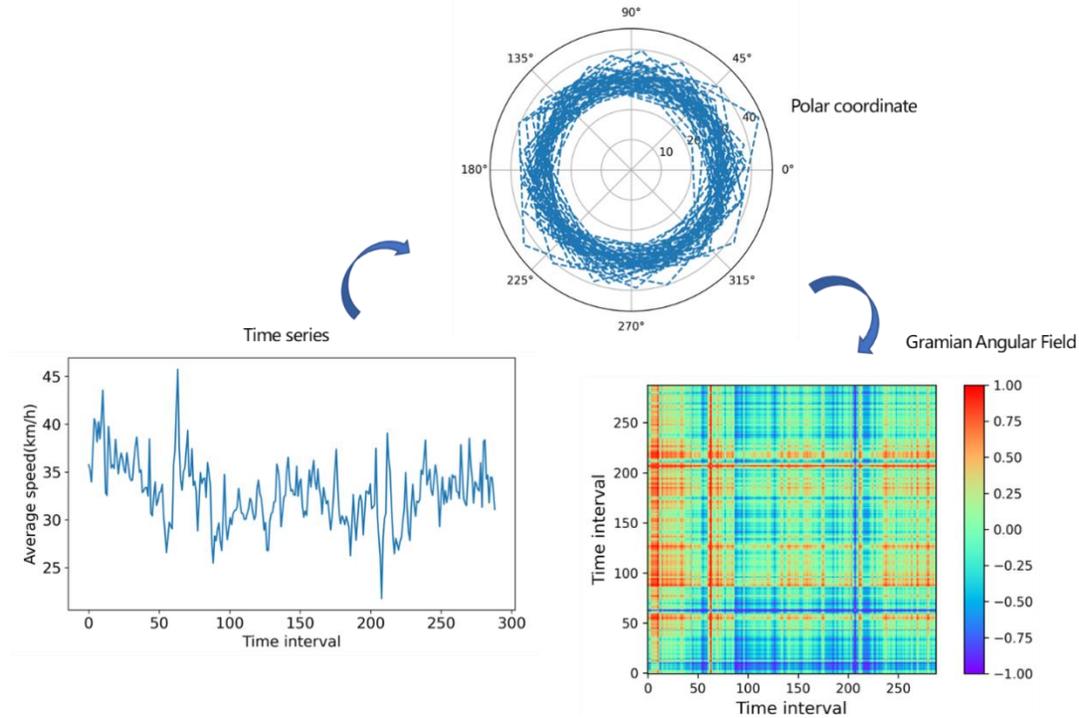

**Figure 2 Illustration of the encoding map of Gramian Angular Field**

### 4.2 Extracting operation features using convolutional autoencoder

After obtaining the daily GAF of all roads, a convolutional autoencoder is built to compress data and extract features. Autoencoder (AE) is a kind of unsupervised learning and representation learning method, which is composed of an encoder and decoder. The working principle of AE is to map the original data to a latent feature space through the encoder (usually the dimension of this space is much smaller than the original data), and then use the decoder to remap it to the original data. Encoder and decoder are usually implemented by the neural network structure. Its main applications include data compression, data denoising, feature extraction, and so on. The goal of AE is to minimize reconstruction error, and its objective function is as follows:

$$\mathcal{L} = \sum_{n=1}^{N} ||x^{(n)} - g(f(x^{(n)}))||^2 \tag{4}$$

$x^{(n)}$ is the original data; $f(x^{(n)})$ is the encoder output; $g(f(x^{(n)}))$ is the decoder output; $\mathcal{L}$ is the sum of squared reconstruction error.

The structure of the convolutional AE in this research is shown in Figure 3, mainly built by some convolutional layers, and both the input and output of this neural network are GAFs. During the reconstruction process of data, the AE learns the latent characteristics in its hidden layers. The encoder part outputs the compressed and denoised data. Since the shape of convolution filter is 3x3 and the stride is 2 in every layer, the dimension of the output of encoder is 9x9x128 as Figure 3 shows, which are length, width, and channels dimensions respectively. We average the outputs in the channel dimension and then flatten them so that each time series is represented as a vector of length 81, whose length is 288 originally.

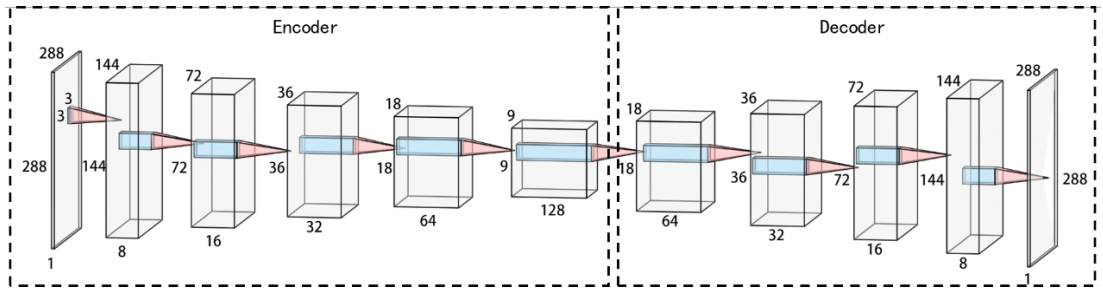

**Figure 3 CNN autoencoder used for features extraction**

### 4.3 Spatial clustering based on road network topology and operation characteristics

The law of spatial correlation proposed by Waldo (*26*) points out that "All things are related, but nearby things are more related than distant things". The correlation between things is related to distance generally. In most cases, the closer the distance, the greater the correlation between things; the farther the distance, the greater the difference between things. Transportation networks have strong spatial correlations due to their topology characteristics, which are clearly reflected in the congestion propagation process (*3*). If a link of the road network is congested at some time points, then the adjacent links have a higher probability to be congested at the same time or get congested soon than far links.

Given this inspiration, on the basis of general hierarchical clustering method, the spatial adjacency relationship of roads is considered. Specifically, only adjacent roads can be in the same partition using spatial hierarchical clustering. The adjacency relationship is shown in Figure4 where three two-way roads cross. When an intersection is formed, roads are considered adjacent. The schematic graph of the clustering process in road network is shown in Figure 5. The clustering vector is the encoder output $f(x^{(n)})$ for all roads. The adjacency relationship is regarded by the adjacent matrix which is a 0-1 matrix.

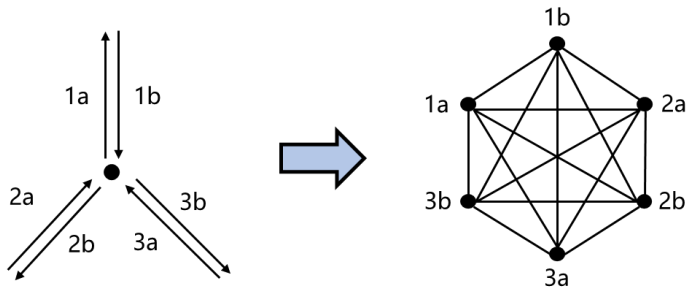

**Figure 4 Spatial adjacency relationship**

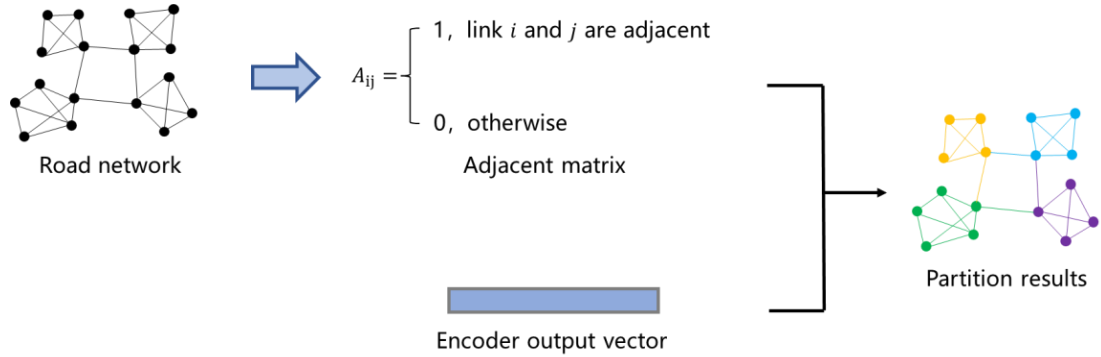

**Figure 5 Spatial hierarchical clustering process**

The idea can be better explained with a case where some heterogenous congested roads exist in the network. The spatial hierarchical clustering method continuously merges adjacent roads with similar congested levels. We can conclude that the roads in the same cluster will show similar behaviors. The significances of hierarchical clustering considering adjacency are as follows: (i) Congestion in an area can only spread to other areas through adjacent roads and usually weakens with the increase of distance. Spatial hierarchical clustering makes full use of this feature; (ii) Sub-networks with internally connected roads make it easier to implement perimeter or regional traffic control strategies. For example, a common way of regional traffic control is to adjust the signal timing, which required the cooperation of many intersections.

## 5.   RESULTS
### 5.1 Shenzhen urban road network partitioning

The results for different numbers of sub-networks using AE-hierarchical clustering are shown in Figure 6. The results have some interesting features as follows:

(i) The tidal traffic characteristic of some two-way roads is distinguished by our method. Affected by the distribution of workplaces and residential areas, there are tidal traffic phenomena on some roads. During morning peak hours, the direction to workplaces is usually congested, while the other direction is contrary. During evening peaks, the situation is just the opposite. These two directions are more likely to belong to different sub-networks.

(ii) The shape of the sub-networks is irregular, and the roads far away in space may be in the same sub-network. Since the generation and dissipation of congestion are spread through adjacent roads, the results prove the ability of our method in finding state-similar roads.

(iii) The shape of sub-networks is related to the administrative regions in Figure 1, but not exactly the same. Though administrative regions affect the spatiotemporal characteristics of road network traffic, regional traffic management and control completely based on administrative regions is clearly not the best option.

The above findings reflect the effectiveness of the partitioning method proposed in this study to some extent. To further evaluate its performance, quantitative indicators and comparisons with other algorithms are needed.

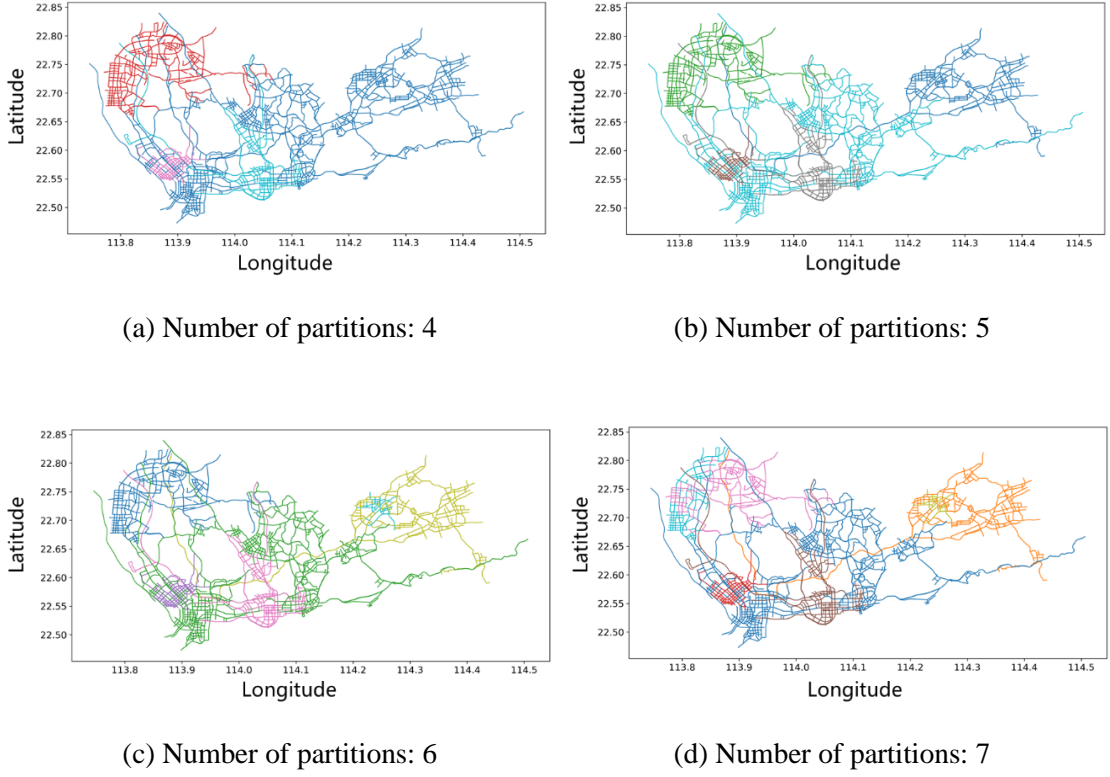

| (a) Number of partitions: 4 | (b) Number of partitions: 5 |
|---|---|
| (c) Number of partitions: 6 | (d) Number of partitions: 7 |

**Figure 6 Partitioning results for Shenzhen network**

### 5.2 Evaluation and comparison of results

To evaluate our framework from different perspectives, the intra and inter metrics are referenced (*4*) and modified based on our method. Intra partition homogeneity is calculated by the intra metric defined in Equation 5. For each sub-network, the average absolute distance is computed between the pair of nodes, and the average of that computed for all the sub-networks is taken. $v_p$ and $v_f$ are the daily average speed series of link $p$ and link $f$. Lower intra value means better results.

$$Intra(P) = \frac{1}{|P|} \sum_{P_i \in P} \frac{\sum_{p \neq q}^{v_p, v_q \in P_i} \sum abs(v_p - v_f)}{|P_i| \cdot (|P_i| - 1)} \tag{5}$$

Inter partition heterogeneity is calculated by the inter metric defined in Equation 6. $P_i \overleftrightarrow{adj} P_j$ denotes the adjacency relation between sub-networks. Inter value is the average absolute distance between each pair of spatial adjacent sub-networks. Higher intra value means better results.

$$Inter(P) = \frac{1}{\left|P_i \overleftrightarrow{adj} P_j\right|} \sum_{P_i, P_j \in P; \, P_i \overrightarrow{adj} P_j} \frac{\sum_{v_p \in P_i} \sum_{v_q \in P_j} \sum abs(v_p - v_f)}{|P_i| \cdot |P_j|} \tag{6}$$

Furthermore, to better demonstrate the advantage of the proposed partition framework, spectral clustering method is utilized to make a comparison. Spectral clustering is a classic graph-partition algorithm, which is widely used in community detection (*11*), computer processing unit allocation (*10*), and many other fields. Since the similarity matrix of roads is needed in the spectral clustering method, the Euclidean distance between series is used to represent the similarity between two links which is defined in Equation 7. To consider the adjacency relationship in spectral clustering, their similarity is not zero only when link $l_p$ and link $l_f$ are adjacent.

$$Sim(l_p, l_f) = \begin{cases} \dfrac{1}{\sqrt{\sum(v_p - v_f)^2} + 0.1}, if\ l_p \overleftrightarrow{adj} l_f \\ 0, otherwise \end{cases} \tag{7}$$

Comparing the results of spectral clustering with AE-hierarchical, we can clearly see that the results in Figure 7 are very similar to administrative regions in Figure 1. This is because spectral clustering tends to cut the graph with weak connections, while Shenzhen has fewer roads between administrative regions.

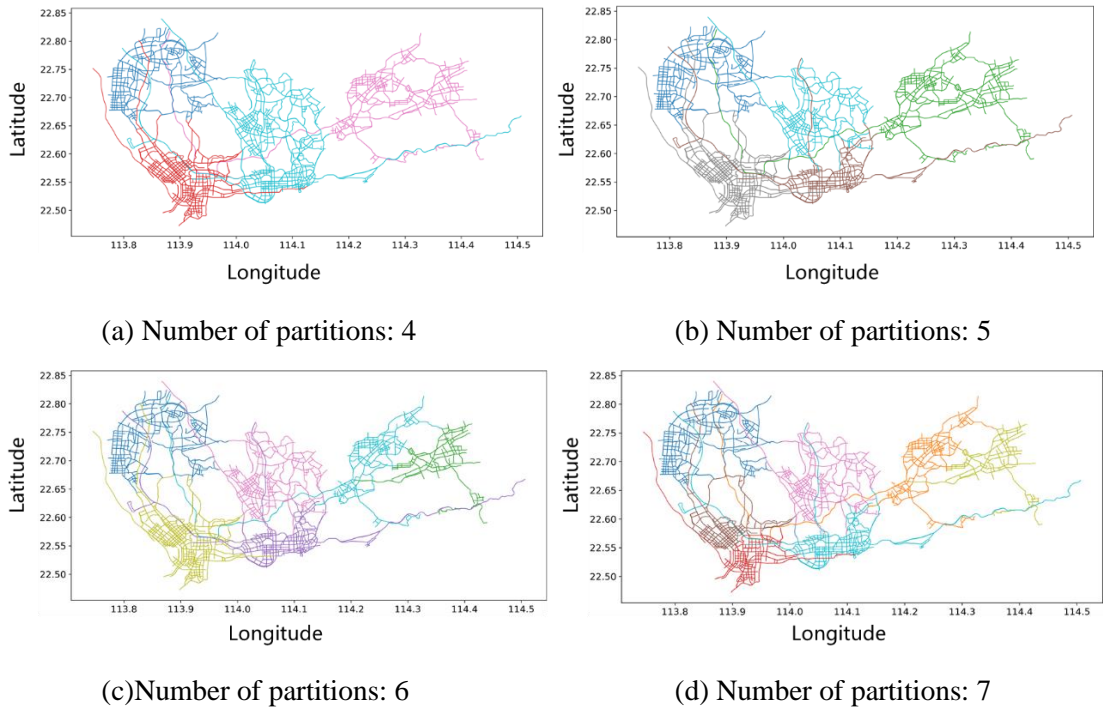

(a) Number of partitions: 4          (b) Number of partitions: 5

(c)Number of partitions: 6          (d) Number of partitions: 7

**Figure 7 Partitioning results for Shenzhen network using spectral clustering**

The intra and inter values of the two methods under a different number of sub-networks are shown in Figure 8 and Figure 9. On average, the former increases by about 9%, and the latter increases by about 9.5% through our methods. For the intra metric, the AE-Hierarchical clustering achieves lower values under all numbers of partitions and the minimum is reached when there are 4 partitions. For the inter metric, the AE-Hierarchical clustering achieves higher values in most cases (except 6, 7 ,8 and 9 partitions) and the maximum is reached when there are only 2 partitions. The possible reason for this situation is that spatial hierarchical clustering continues to absorb adjacent roads with similar traffic states to form clusters. In this process, roads with similar states are more likely to enter the same cluster. However, due to its priority to ensure intra homogeneity, it will inevitably lead to the weakening trend of inter heterogeneity as a whole when the number of partitions increases. We also compute the intra homogeneity of the whole road network to explain the partitioning impact.

$$Intra(M) = \frac{\sum_{p \neq q}^{v_p, v_q \in M} \sum abs(v_p - v_f)}{|M| \cdot (|M| - 1)} \qquad (8)$$

Where *M* refers to the whole road network. The calculation result is much large (>1000), which reflects the strong heterogeneity of the traffic state in the road network and the effectiveness of partitioning methods.

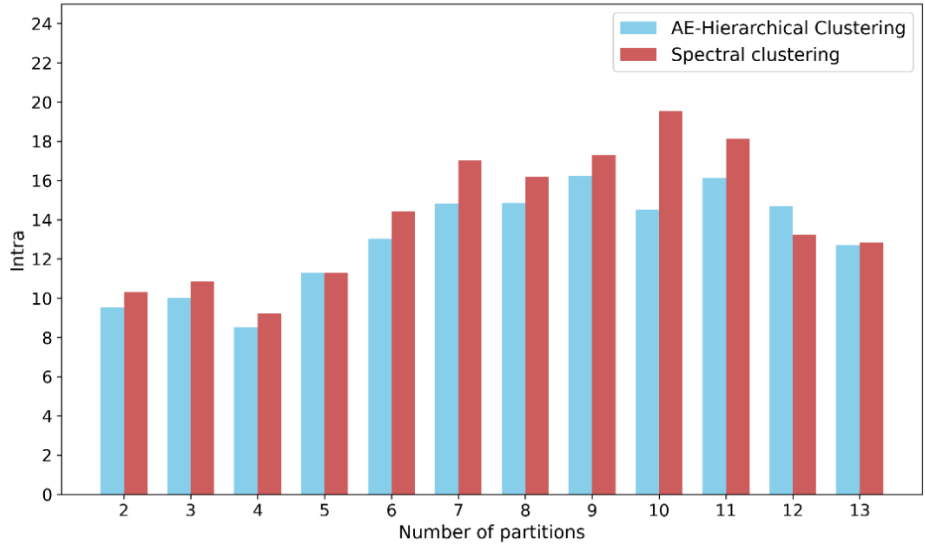

**Figure 8** Intra homogeneity of the two methods under different number of partitions

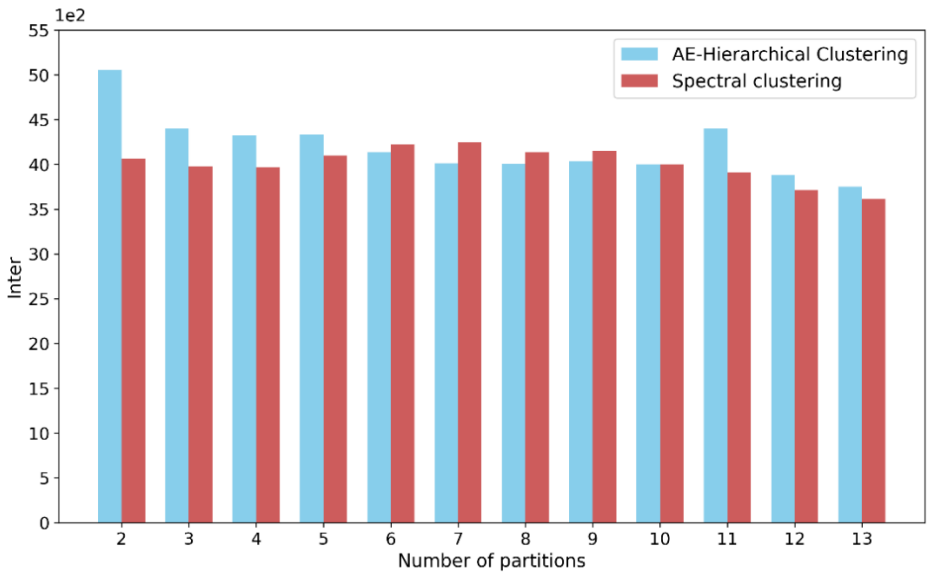

**Figure 9** Inter heterogeneity of the two methods under different number of partitions

Furthermore, the time-consuming partitioning is compared. Although the training of the convolutional autoencoder takes a relatively long time, it is sufficient to retrain the model periodically or when the road network topology changes significantly. Therefore, we only calculate the time when the encoder receives the input and obtains the output plus the time for spatial hierarchical clustering. Spectral clustering only includes the operation time of the partition process, excluding the calculation time of the similarity matrix. The result is shown in Figure 10. Spectral clustering is much more time-consuming than the AE-Hierarchical clustering proposed in this study.

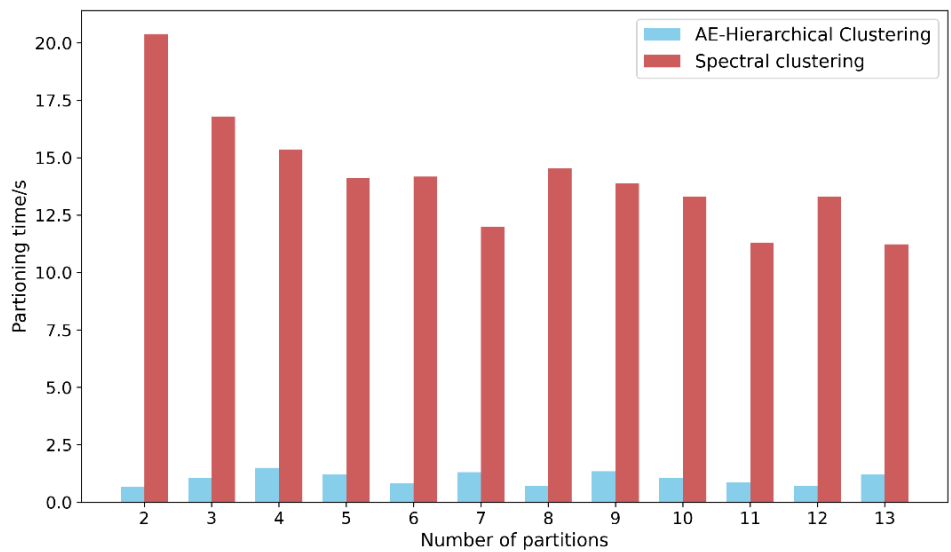

**Figure 10 Partitioning speed of the two methods**

## 6. DISCUSSION

Urban road traffic system has a high degree of spatio-temporal heterogeneity. It is necessary to identify different clusters depending on the distribution of traffic states. The propagation of congestion is highly related to heterogeneity and is usually considered when partitioning. This study proposes a complete methodology for the urban heterogeneous road network partitioning using deep learning methods.

Recent studies mostly partition based on the short-term traffic state of the road section (*3–5*, *9*, *18*), but the disadvantage is that the sub-networks change quickly (for example, partition every five minutes), which requires high collaborative response ability of software and hardware. Meanwhile, unstable partitions are difficult to cooperate with traffic management and control measures. This study comprehensively considers the road traffic status of a day. Although this will lose a small amount of accuracy, it also reduces the difficulty of model application, and can better cooperate with traffic guidance and other measures.

Partitioning makes perimeter control strategies easier through the Macroscopic Fundamental Diagram (MFD). MFD is a well-defined curve existing between space-mean flow and density at network level. The idea of MFD was first proposed by Godfrey (*27*) and re-introduced by Mahmassani and Chang (*28*), Herman and Prigogine (*27*), Daganzo and Geroliminis (*30*, *31*). Some studies (*30–32*) showed that the main factor affecting MFD is the spatial distribution of vehicle density in the road network. These findings are very important since the MFD can be applied for heterogeneous road networks, if the whole network can be partitioned into some homogeneous sub-networks. Through partitioning, we can obtain (i) sub-networks with small variance, which increases the network flow for the same average density (*3*); (ii) spatially connected and compact sub-networks which makes the perimeter control feasible and precise. The proposed mechanism in this study can generate a partition with required number of sub-networks, which contains connected roads with small state variance.

Another interesting research direction is to explore the role of partitioning on trunk control and region control. Trunk control hopes to find key components with large traffic flow in the road network for coordination control, and the proposed methodology based on congestion propagation characteristics can find out these key components. For region control, homogeneous regions are less difficult and more effective to implement related measures. For example, regional intersection signal control system SCATs and SCOOT are widely used in China metropolises, and their architectures are mostly based on administrative regions. There may be a better control effect if they are based on homogeneous sub-networks.

## CONCLUSIONS

In this study, a framework is proposed to partition the urban road network. The proposed methodology includes the time series encoding to the matrix, the characteristics extraction using the convolutional autoencoder, and the spatial hierarchical clustering based on the encoder output. An empirical study of Shenzhen is conducted to demonstrate the effectiveness of the framework. To quantitatively evaluate the partitioning results, the classic spectral clustering method is chosen as a baseline. The intra homogeneity within each sub-network and inter heterogeneity between sub-networks are calculated under the various number of partitions. The results are summarized as below:

(i) Compared with the classic graph partition method, the proposed framework has advantages. There is a 9% and 9.5% increase of the intra homogeneity and inter heterogeneity for obtained sub-networks respectively. There is also a great reduction in the running time.

(ii) Partitioning based on congestion propagation characteristics is an effective and useful idea in terms of results. Since congestion spreads through adjacent roads, it will naturally form regional homogeneous connected road clusters, which is also in line with the purpose of regional traffic control. From our results, the shape of sub-networks is irregular, which is an embodiment of congestion propagation

(iii) Tidal traffic is distinguished by our methods. Some two-way roads belong to different sub-networks, which are mainly caused by the separation of work and residence. This study may contribute to the management and control of tidal traffic problems (*33–35*) generally happened in urban road networks.

However, there are still some shortcomings for further improvement in future works. Due to the limitation of data, this study is based on roads rather than road segments, so the results are rough to some extent. In addition, the interpretability of partitioning results needs to be strengthened, which perhaps can be further explained by the temporal and spatial distribution characteristics of road network traffic states before and after partition.

## ACKNOWLEDGMENTS

This work was supported by the National Key R&D Program of China (No. 2018YFB1601100).

## AUTHOR CONTRIBUTIONS

The authors confirm contributions to the paper as follows: study conception and design: Jian Li, Weifeng Li and Pengfei Xu; data collection: Jian Li and Weifeng Li; analysis and interpretation of results: Pengfei Xu, Jian Li and Weifeng Li; draft manuscript preparation: Pengfei Xu, Jian Li, Weifeng Li and Chenjie Xu. All authors reviewed the results and approved the final version of the manuscript.